\documentclass[letterpaper,twocolumn,english]{revtex4-1}
\usepackage[T1]{fontenc}
\usepackage[latin9]{inputenc}
\usepackage{xcolor}
\usepackage{pdfcolmk}
\usepackage{booktabs}
\usepackage{bm}
\usepackage{amsmath}
\usepackage{amssymb}
\usepackage{graphicx}
\PassOptionsToPackage{normalem}{ulem}
\usepackage{ulem}
\usepackage{subscript}

\makeatletter

\pdfpageheight\paperheight
\pdfpagewidth\paperwidth

\providecommand{\tabularnewline}{\\}
\providecolor{lyxadded}{rgb}{0,0,1}
\providecolor{lyxdeleted}{rgb}{1,0,0}

\DeclareRobustCommand{\lyxsout}[1]{\ifx\\#1\else\sout{#1}\fi}

\usepackage{babel}

\makeatother

\usepackage{babel}
\begin{document}
\title{Exciton-polariton mediated interaction between two nitrogen-vacancy
color centers in diamond using two-dimensional transition metal dichalcogenides}
\author{J. C. G. Henriques$^{1}$, B. Amorim$^{1}$, N. M. R. Peres$^{1,2}$}
\address{$^{1}$Department and Centre of Physics, and QuantaLab, University
of Minho, Campus of Gualtar, 4710-057, Braga, Portugal}
\address{$^{2}$International Iberian Nanotechnology Laboratory (INL), Av. Mestre
José Veiga, 4715-330, Braga, Portugal}
\begin{abstract}
In this paper, starting from a quantum master equation, we discuss
the interaction between two negatively charged Nitrogen-vacancy color
centers in diamond via exciton-polaritons propagating in a two-dimensional
transition metal dichalcogenide layer in close proximity to a diamond
crystal. We focus on the optical 1.945 eV transition and model the
Nitrogen-vacancy color centers as two-level (artificial) atoms. We
find that the interaction parameters and the energy levels renormalization
constants are extremely sensitive to the distance of the Nitrogen-vacancy
centers to the transition metal dichalcogenide layer. Analytical expressions
are obtained for the spectrum of the exciton-polaritons and for the
damping constants entering the Lindblad equation. The conditions for
occurrence of exciton mediated superradiance are discussed.
\end{abstract}
\maketitle

\section{Introduction}

\begin{figure}[h]
\centering{}\includegraphics[width=8cm]{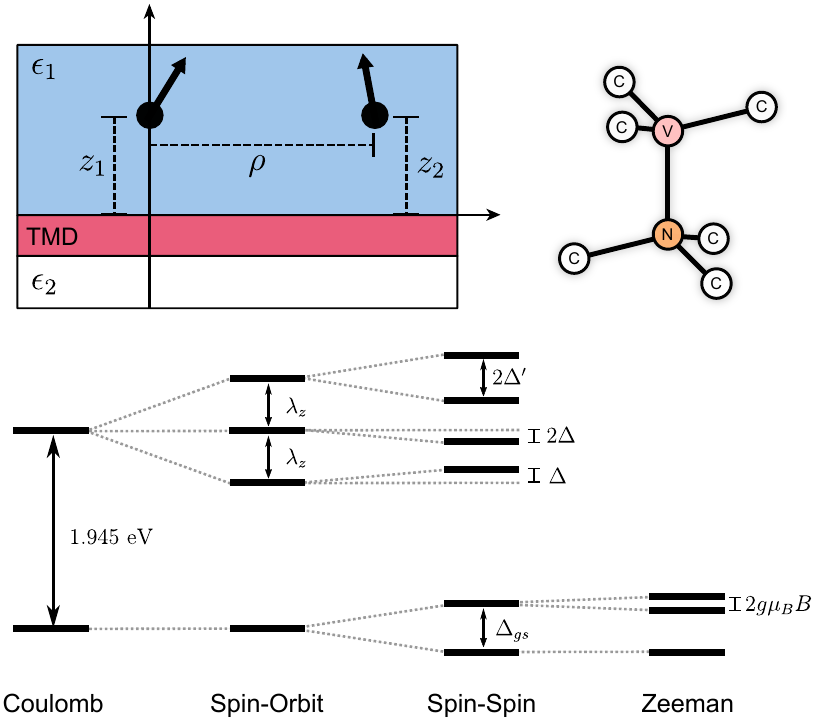}\caption{\label{fig:system}System considered in this paper: Two NV-centers
each carrying an electric dipole moment $\bm{\mu}_{j}$ (arrows),
associated with the optical transition of $1.945$ eV, are embedded
in a diamond slab with a dielectric constant $\epsilon_{1}$. The
slab is on top of a single layer of a TMD, and the whole system is
on top of a dielectric with dielectric function $\epsilon_{2}$ (the
value of $\epsilon_{2}$ allows the tuning of the exciton frequency).
Each NV-center is at a distance $z_{i}$ from the TMD layer and is
separated from one another by a distance $\rho$. On the right panel
we depicted the crystal structure of diamond carrying a NV-center:
nitrogen (orange) and vacancy (pink). In the bottom panel, the energy
diagram of a NV-center is depicted together with the different mechanisms
leading to the energy-levels spiting (adapted from Ref. \citep{Fabre:2017}).
The transition we will be considering is the optical transition of
$1.945$ eV.}
\end{figure}

Nitrogen-vacancy color centers (NV-centers) are fascinating artificial
atoms in diamond, with electronic transitions in the visible spectral
range\citep{Jelezko2006,Hong:2013,Doherty:2013aa}. Nowadays they
can be implanted with atomic precision in nano-diamond-layers as thin
as 5 nm \citep{Ohashi:2013aa}. One type of NV-centers in diamond
is the charged neutral NV$^{0}$ one \citep{Hauf:2011aa,Fu:2010aa}.
There is, however, the possibility of producing negatively charged
NV-centers both chemically \citep{Fu:2010aa,Hauf:2011aa,Doherty:2011aa}
or, more interestingly, using an external gate \citep{Lillie:2019aa}.
This latter possibility brings extra tunability to the charge control
of\textcolor{black}{{} these structures. }The NV-center shows an electron
paramagnetic ground state and an optically excited state. The energy
ground state of the charged NV-center forms a spin triplet (due to
two electrons sitting on the vacancy) which due to spin-spin interaction
is split into\textcolor{black}{{} two }energy levels separated by $\Delta_{gs}=2.87$
GHz \citep{Hong:2013} (see bottom panel of Fig. \ref{fig:system}).
The exploration of these energy levels for quantum optics \citep{Fabre:2017}
and for quantum computation has already been considered \citep{Liu:2013aa}.
In addition to these low energy levels, there is a high-energy state
(energy difference of 1.945 eV; see Fig. \ref{fig:system}) which
is at the origin of the pinkish color in diamonds \citep{Jelezko2006}.
Negatively charged NV-centers also acquire an electric dipole moment
\citep{Jelezko2006}. This has its origin on the localized nature
of the electronic wave-function. The negatively charged NV-centers,
coupled electrically, can act as interacting quantum bits (qubits)
useful in quantum computation \citep{Thiering:2020aa}.

Among the many applications of NV-centers we underline their high
sensing capability of extremely weak magnetic fields with high spatial
resolution \citep{Hong:2013,Schirhagl:2014aa}, their application
in quantum computation \citep{Bernien:2013aa,Sipahigil:2014aa,Thiering:2020aa}
and in quantum nanophotonics \citep{Bradac:2019aa}, and, more recently,
their use as a microscopy tool for characterizing field effect transistors
made of two-dimensional materials \citep{Lillie:2019aa} as well as
in superconducting materials \citep{Acosta:2019aa}. Applications
to metrology have also been demonstrated \citep{Dolde:2014aa}.

It is well known that quantum emitters can interact with each other
via electromagnetic radiation \citep{Lehmberg:1970aa}. When in the
presence of a metallic surface capable of supporting surface-plasmon-polaritons
(SPPs), the interaction between atoms, via this special type of electromagnetic
mode, can be tuned by changing the distance and relative orientation
of the electric dipole moment of the atoms \citep{Gonzalez-Tudela:2011aa,Martin-Cano:2011aa},
and, more importantly, changing the thickness of metallic film in
the vicinity of the diamond layer \citep{Zhou:2017aa,Torma:2014aa,Delga:2014aa}.
Indeed, we can even take the limit of vanishing thickness of the metallic
layer using doped graphene as the plasmonic material in the proximity
of the diamond layer. Graphene supports SPPs in the mid-IR which can
mediate interactions between neighboring atoms (or artificial atoms
for the same matter) \citep{Huidobro:2012aa}. Surface plasmons polaritons
are not the only existing kind of polaritonic surface waves. Other
types of this kind of waves are phonon-polaritons, propagating at
the surface of polar dielectric materials, such as hexagonal boron
nitride \citep{Chaudhary:2019aa} or silicon oxide \citep{Chen:2007aa}.

Recently, a new kind of surface polariton has been proposed in Ref.~\citep{Epstein:2020aa}:
exciton-polaritons supported, at low temperatures, by two-dimensional
semiconducting transition metal dichalcogenides (TMDs) - MX$_{2}$
with M = Mo, W and X = S, Se, Te. These polaritons, as in the case
of SPPs, form when the real-part of the dielectric function is negative.
This happens in a small energy window occurring close to the bare
exciton energy, in the visible spectral range. The energy levels of
negatively charged NV-centers can also occur in the same spectral
range (they are color centers). Therefore, it is conceivable that
two negatively charged NV-centers, when positioned in close proximity
to a two-dimensional layer of a TMD, can strongly interact via exciton-polaritons.
In addition to the aforementioned interaction, the energy levels of
the NV-centers are also renormalized by the interaction mediated by
the polaritons. \uwave{Also, spin-spin interactions among distant
NV-centers may also be mediated by exciton-polaritons \citep{Quinteiro:2006aa}}\textcolor{red}{\uwave{.}}

In this work, we will derive a quantum master equation, also termed
optical Lindblad equation in the context of quantum optics \citep{Brasill:2013,Schaller:2014},
to study the physics of NV-centers coupled to TMD exciton-polaritons.
The coupling to exciton-polaritons generates couplings between the
different NV-centers, renormalization of NV-centers energy transitions
and also open collective decay paths.The structure of the paper is
the following: In Sec. \ref{sec:Hamiltonian}, we present the model
Hamiltonian used to describe the coupled NV-center/exciton-polariton,
as well as the quantum master equation that governs the NV-centers,
and express the NV-centers transition energy shifts, exction-polation
mediated couplings and dissipators in terms of exciton-polariton Green's
functions. In Sec. \ref{sec:Results}, we discuss how these different
parameters depend on the separation between the NV-centers and on
their dipole orientation\textcolor{black}{s. We also analyze the possibility
of observing superradiance mediated by the exciton-polaritons. }In
Sec \ref{sec:Conclusions}, we summarize the main conclusions of the
work. \textcolor{black}{An appendix with some derivation calculation
details is present at the end of manuscript.}

\section{Model\label{sec:Hamiltonian}}

We consider that a TMD monolayer, supporting exciton-polaritons, is
located at the $z=0$ plane. The optical conductivity of the two-dimensional
TMD is denoted by $\sigma(\omega)$. The NV-centers are above the
TMD layer, $z>0$. We describe the NV-centers as two-level systems
embedded in a dielectric medium with constant $\epsilon_{1}$, for
which we take the dielectric constant of diamond. The two NV-centers
are located at positions $\mathbf{r}_{i}$, with the index $i=1,2$
labeling the NV-center. The TMD has underneath, $z<0$, a different
dielectric medium, with dielectric constant $\epsilon_{2}$, which
we take as vacuum. The system is depicted in Fig.~\ref{fig:system}.

\subsection{Hamiltonian}

We model the coupled system of TMD exciton-polartions and NV-centers
with the\textcolor{black}{{} }modified version of the Dicke model where
the NV-center two-level systems are coupled to a multimode boson field,
the exciton-polarion field \cite{Garraway:2011aa,Cong:2016,Kirton:2019aa,Cortes:2020aa}.
Explicitly we have the Hamiltonian

\begin{equation}
H=H_{\text{NV}}+H_{\text{ex-p}}+H_{\text{int}},\label{eq:ModelHamiltonian}
\end{equation}
where $H_{\text{NV}}$, $H_{\text{ex-p}}$, and $H_{\text{int}}$
are, respectively, the Hamiltonian for the NV-centers, TMD exciton-polaritons,
and NV-centers/exciton-polariton interaction. We model the NV-centers
as two-level systems, such that $H_{\text{NV}}$ reads: 
\begin{equation}
H_{\text{NV}}=\sum_{i=1}^{2}\frac{1}{2}\hbar\omega_{0}\sigma_{i}^{z},
\end{equation}
with $\hbar\omega_{0}$ the energy difference between the two energy
levels of the NV-centers, $\sigma_{i}^{z}$ the $z$ Pauli matrix,
written in the basis $\left\{ \left|e\right\rangle _{i},\left|g\right\rangle _{i}\right\} $,
where $e$ and $g$ refer, respectively, to the excited and ground
state of NV-center $i$. The TMD exciton-polaritons are modeled as
independent bosons governed by: 
\begin{equation}
H_{\text{ex-p}}=\sum_{\mathbf{q}}\hbar\omega_{\mathbf{q}}a_{\mathbf{q}}^{\dagger}a_{\mathbf{q}},
\end{equation}
where $\hbar\omega_{\mathbf{q}}$ corresponds to the energy dispersion
relation of the exciton-polaritons with momentum $\mathbf{q}$,\textcolor{red}{{}
}\textcolor{black}{whose analytical expression will be given later},
and $a_{\mathbf{q}}^{\dagger}$ ($a_{\mathbf{q}}$) corresponds to
the creation (annihilation) of an exciton-polariton of momentum $\mathbf{q}$.
Finally, the interaction between exciton-polartions and NV-centers
is modeled in the dipole coupling approximation and reads: 
\begin{equation}
H_{\text{int}}=\sum_{i=1}^{2}\mathbf{E}_{\text{ex-p}}(\mathbf{r}_{i})\cdot\left(\boldsymbol{\mu}_{i}\sigma_{i}^{-}+\bm{\mu}_{i}^{*}\sigma_{i}^{+}\right),
\end{equation}
where $\boldsymbol{\mu}_{i}$ the electric dipole moment (associated
with the optical transition and whose absolute maximum value is about
1.5 D, as measured from the Stark shift \cite{Tamarat:2006aa,comment};
see also \cite{Gali:2019aa}) of the NV-center $i$, and $\sigma_{i}^{\pm}$
are raising/lowering operators, represented in the $\left\{ \left|e\right\rangle _{i},\left|g\right\rangle _{i}\right\} $
basis as 
\begin{equation}
\sigma_{i}^{+}=\left(\begin{array}{cc}
0 & 1\\
0 & 0
\end{array}\right),\quad\sigma_{i}^{-}=\left(\begin{array}{cc}
0 & 0\\
1 & 0
\end{array}\right),
\end{equation}
\textcolor{black}{{} $\mathbf{E}_{\text{ex-p}}(\mathbf{r})$ is the
exciton-polariton eletric field operator, which is written as
\begin{equation}
\mathbf{E}_{\text{ex-p}}(\mathbf{r})=\sum_{\mathbf{q}}\left(\mathbf{E}_{\mathbf{q},\text{ex-p}}(\mathbf{r})a_{\mathbf{q}}+\mathbf{E}_{\mathbf{q},\text{ex-p}}^{*}(\mathbf{r})a_{\mathbf{q}}^{\dagger}\right),\label{eq:E_expansion}
\end{equation}
with $\mathbf{E}_{\mathbf{q},\text{ex-p}}(\mathbf{r})$ the exciton-polariton
electric field mode function. Following \cite{Ferreira:2020aa}, the
exciton-polariton mode function for the considered structure is given
by} \citep{Ferreira:2020aa,You:2020aa}: 
\begin{align}
\mathbf{E}_{\mathbf{q},\text{ex-p}}(\mathbf{r}) & =i\sqrt{\frac{\hbar\omega_{\mathbf{q}}}{2A\epsilon_{0}N_{\mathbf{q}}}}e^{i\mathbf{q}\cdot\mathbf{x}}\nonumber \\
 & \cdot\begin{cases}
\left(i\frac{\mathbf{q}}{q}-\frac{q}{\kappa_{1,\mathbf{q}}}\hat{\mathbf{z}}\right)e^{-\kappa_{1,\mathbf{q}}z}, & z>0\\
\left(i\frac{\mathbf{q}}{q}+\frac{q}{\kappa_{2,\mathbf{q}}}\hat{\mathbf{z}}\right)e^{\kappa_{2,\mathbf{q}}z}, & z<0
\end{cases},\label{eq:u(q) mode function}
\end{align}
where $\mathbf{r}=(\mathbf{x},z)$, with $\mathbf{x}$ an in-plane
two dimensional vector, $A$ is the area of the system, $\epsilon_{0}$
is the electric permittivity of vacuum, $\kappa_{n,\mathbf{q}}$ is
an out of plane momentum defined as: 
\begin{equation}
\kappa_{n,\mathbf{q}}=\sqrt{q^{2}-\omega_{\mathbf{q}}^{2}\epsilon_{n}/c^{2}},
\end{equation}
with $c$ the speed of light and $\epsilon_{n}$ the dielectric constant
of the medium; $N_{\mathbf{q}}$ is a mode length, that originates
from the normalization of the mode in a dispersive medium. Its exact
form is provided in Appendix~\ref{appx:Mode-length-and}.

\subsection{Exciton-polariton dispersion relation}

We will now specify the dispersion relation of the exciton-polaritons.
This depends on the optical properties of a monalayer semiconducting
TMD, which are described in terms of its optical conductivity, $\sigma(\omega)$.
The optical conductivity is related to the susceptibility, $\chi(\omega)$,
via the relation \cite{Moradi:2020}
\begin{equation}
\sigma(\omega)=-id\omega\epsilon_{0}\chi(\omega).\label{eq:sigma(omega)}
\end{equation}
where $d$ is the monolayer thickness and $\epsilon_{0}$ is the vacuum
permittivity. At low temperatures and for frequencies close to the
exciton, the susceptibility is accurately modeled by the formula \cite{Epstein:2020aa}:

\begin{equation}
\chi(\omega)=\chi_{\text{bg}}-f_{\text{ex}}\frac{\omega_{\text{ex}}^{2}}{\omega^{2}-\omega_{\text{ex}}^{2}+i\omega\left(\frac{\gamma_{\text{nr}}}{2}+\gamma_{\text{d}}\right)},\label{eq:chi(=00003D00005Comega)}
\end{equation}
where $\omega_{\text{ex}}$ is the exciton's energy, $f_{\text{ex}}$
is the exciton oscillator strength, which describes the coupling of
the exciton to the electric field, $\gamma_{\text{nr}}$ and $\gamma_{\text{d}}$
are, respectively, the non-radiative and dephasing decay rates, and
$\chi_{\text{bg}}$ is a background contribution to the susceptibility
(which takes into account higher energy transitions). Notice that
all this parameters are device dependent. In Table~\ref{tab:parameters},
we report typical values which we will use hereinafter. For positive
frequencies close to $\omega_{\text{ex}}$ and weak losses ($\gamma_{\text{nr}},\gamma_{\text{d}}\ll\omega_{\text{ex}}$)
the susceptibility is generally approximated by
\begin{equation}
\chi(\omega)\simeq\chi_{\text{bg}}-\frac{1}{2}f_{\text{ex}}\frac{\omega_{\text{ex}}}{\omega-\omega_{\text{ex}}+i\left(\frac{\gamma_{\text{nr}}}{2}+\gamma_{\text{d}}\right)}.
\end{equation}

\begin{table*}
\begin{centering}
\begin{tabular}{cccccccc}
\toprule 
$\hbar\omega_{0}$ (eV) & $\mu$ (D) & $\hbar\omega_{\text{ex}}$ (eV) & $f_{\text{ex}}$ & $\gamma_{\text{nr}}$ (meV) & $\gamma_{\text{d}}$ (meV) & $d$ (nm) & $\chi_{\text{bg}}$\tabularnewline
\midrule
\midrule 
1.945 & 1.5 & 1.94 & 0.39 & 1.99 & 0.04 & 0.65 & 15\tabularnewline
\bottomrule
\end{tabular}
\par\end{centering}
\centering{}\caption{\label{tab:parameters}Parameters used throughout this work. The energy
$\hbar\omega_{0}$ corresponds to the energy difference between the
NV-center levels (Ref. \cite{Tamarat:2006aa}). The parameter $\mu$
(Ref. \cite{Tamarat:2006aa}) corresponds to the magnitude of the
electric dipole moment (which we consider to be the same for both
dipoles). $\hbar\omega_{\text{ex}}$ corresponds to a typical energy
of the A-exciton energy in a TMD (Ref. \cite{Wang2018Colloquium}
reports a similar value for WS\protect\textsubscript{2} on SiO\protect\textsubscript{2}
for a temperature of $T=4$ K). $d$ is the effective thickness of
a monolayer TMD (Ref. \cite{Molina2011}). The parameter $\chi_{\text{bg}}$
corresponds to the background contribution to the susceptibility,
that we considered as a free parameter whose value was chosen in order
to observe the desired phenomena. The values for the oscillator strength,
$f_{\text{ex}}$, the non-radiative decay rate, $\gamma_{\text{nr}},$
and dephasing rate , $\gamma_{\text{d}}$, refer to $\text{WS}{}_{2}$
and were taken from \cite{Epstein:2020aa}, for a temperature of $T=4$
K.\label{tab:Parameters-used-throughout}}
\end{table*}

\begin{figure}[h]
\centering{}\includegraphics[scale=0.95]{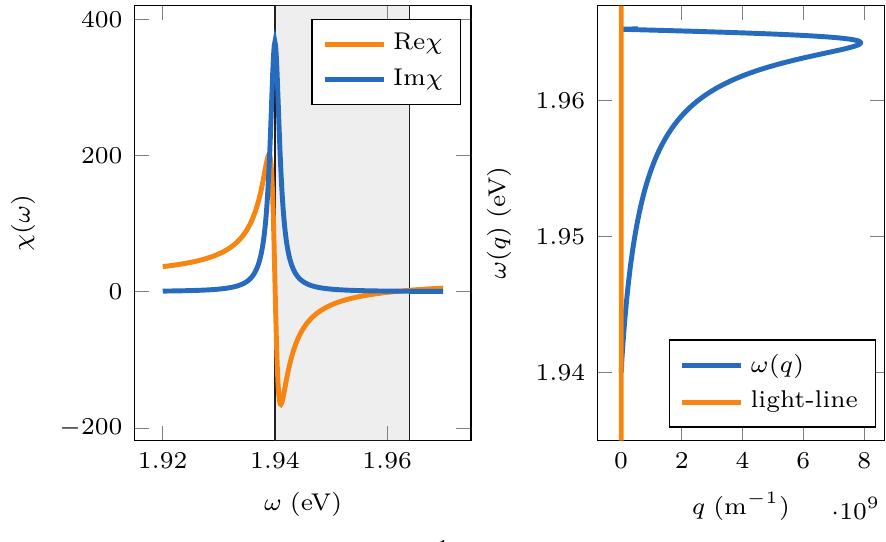}\caption{\label{fig:susceptibility and polariton disp} (Left) Real and imaginary
parts of the susceptibility according to Eq. (\ref{eq:chi(=00003D00005Comega)}).
We observe a resonance at the exciton energy $\omega_{\text{ex}}=1.94$
eV. The real part of the susceptibility is negative in a small interval
from approximately 1.940 eV up to 1.965 eV (shaded region). Only in
this spectral region does the monolayer TMD support exciton-polaritons.
(Right) Exciton-polariton dispersion relation $\omega_{\mathbf{q}}$
using the approximate expression given in Eq. (\ref{eq:approx q(w)})
and the susceptibility presented in the left panel. The spectrum is
only finite in the energy window where $\textrm{Re}\chi$ is negative.
The polariton's momentum is much bigger than that of a photon for
a given energy, which agrees with the argument used prior to Eq. (\ref{eq:approx q(w)}).
The parameters of Table \ref{tab:parameters} were used.}
\end{figure}

For the considered structure, Fig.~\ref{fig:system}, the condition
that determines the exciton-polariton energy dispersion is \citep{Goncalves:2015aa,Ferreira:2019aa}:
\begin{equation}
\frac{\epsilon_{1}}{\kappa_{1,\mathbf{q}}}+\frac{\epsilon_{2}}{\kappa_{2,\mathbf{q}}}+i\frac{\sigma(\omega)}{\epsilon_{0}\omega}=0,\label{eq:polariton spectrum 1}
\end{equation}
where, once again, $\epsilon_{1}$ and $\epsilon_{2}$ the media above
and below the TMD layer, respectively, and $\sigma(\omega)$ is the
optical conductivity of the TMD. Inserting Eq. (\ref{eq:sigma(omega)})
into Eq. (\ref{eq:polariton spectrum 1}), one clearly sees that a
solution for the exciton-polariton dispersion is only possible when
the real part of the susceptibility is negative. Considering the parameters
given in Table \ref{tab:Parameters-used-throughout}, we obtain the
susceptibility depicted in Fig. \ref{fig:susceptibility and polariton disp}.
The energy window where exciton-polaritons may be excited, for the
chosen parameters, is 1.940 eV up to 1.965 eV. Within this energy
range the polaritons will present an energy dispersion relation defined
implicitly by Eq. (\ref{eq:polariton spectrum 1}). When the media
above and below the TMD are the same it is possible to obtain an exact
analytical solution for the energy dispersion. However, when they
are not, the problem can only be solved exactly numerically. If we
consider\textcolor{red}{{} }$\kappa_{1,\mathbf{q}}\simeq\kappa_{2,\mathbf{q}}\simeq q$,
that is, if we consider the polariton's momentum to be much bigger
than that of a photon with the same energy, a condition easily fulfilled,
an approximate analytical solution for $q(\omega)$ can be found:\textcolor{black}{
\begin{align}
q(\omega) & {\color{black}{\normalcolor \approx-\frac{\epsilon_{1}+\epsilon_{2}}{d\chi(\omega)}}}\label{eq:approx q(w)}
\end{align}
}In Fig. \ref{fig:susceptibility and polariton disp} we present the
exciton-polariton spectrum obtained with this expression. We observe
that the spectrum is only defined in the small spectral window where
the real part of the susceptibility is negative. We further note that
although the spectrum is defined over a small range of energies, it
covers a wide range of momenta. \textcolor{black}{From the comparison
of the polariton spectrum with the light-line}, we observe that the
polariton momentum is significantly larger than that of a photon with
the same energy, justifying the approximation made in Eq. (\ref{eq:approx q(w)}).

\section{Lindblad equation\label{sec:Results}}

In this section we will determine the Linblad equation for the NV-centers
and express the different terms that enter it in terms of exciton-polariton
Green's function.

Tracing out the exciton-polariton degrees of freedom, we obtain the
Lindblad equation for the reduced density matrix for the NV-centers
in the Schrödinger picture (details of the derivation are provided
in Appendix~\eqref{appx:Lindblad}): 
\begin{align}
\dot{\rho}(t) & =-\frac{i}{\hbar}\left[\sum_{i}\frac{1}{2}\left(\hbar\omega_{0}+\Delta_{i}\right)\sigma_{i}^{z}+\sum_{i\neq j}g_{ij}\sigma_{i}^{+}\sigma_{j}^{-},\rho(t)\right]\nonumber \\
 & +\frac{1}{\hbar}\sum_{i,j}\gamma_{ij}\left(\sigma_{j}^{-}\rho(t)\sigma_{i}^{+}-\frac{1}{2}\left\{ \sigma_{i}^{+}\sigma_{j}^{-},\rho(t)\right\} \right)\nonumber \\
 & +\frac{1}{\hbar}\sum_{i,j}\tilde{\gamma}_{ij}\left(\sigma_{j}^{+}\rho(t)\sigma_{i}^{-}-\frac{1}{2}\left\{ \sigma_{i}^{-}\sigma_{j}^{+},\rho(t)\right\} \right).\label{eq:Lindblad_equation}
\end{align}
In the above equation $\Delta_{i}$ is the so called Lamb shift, representing
the correction to the transition energy of the NV-centers, $g_{ij}$
are couplings between the NV-centers mediated by the exciton-polaritons,
$\gamma_{ij}$ and $\tilde{\gamma}_{ij}$ are dissipation coefficients.
For a bath in thermal equilibrium and using the quantum fluctuation-dissipation
theorem it is possible to express all these quantities in terms of
the Green's functions of the exciton-polaritons. The electric field
retarded/advanced Green's function for exciton-polaritons is defined
as
\begin{equation}
D_{\alpha\beta}^{R/A}(t;\mathbf{r}_{i},\mathbf{r}_{j})=\mp\frac{i}{\hbar}\Theta\left(\pm t\right)\left\langle \left[E_{\text{ex-p}}^{\alpha}(t,\mathbf{r}_{i}),E_{\text{ex-p}}^{\beta}(0,\mathbf{r}_{j})\right]\right\rangle _{R}.
\end{equation}
where $\left\langle ...\right\rangle _{R}$ represents the quantum
thermodynamic average over the isolated reservoir degrees of freedom
of the bath, and greek indices run over spatial coordinates. Making
a Fourier transform in time, $D_{\alpha\beta}^{R/A}(\omega;\mathbf{r}_{i},\mathbf{r}_{j})=\int dte^{i\omega t}D_{\alpha\beta}^{R/A}(t;\mathbf{r}_{i},\mathbf{r}_{j})$,
we obtain
\begin{equation}
D_{\alpha\beta}^{R/A}(\omega;\mathbf{r}_{i},\mathbf{r}_{j})=\sum_{\mathbf{q}}\frac{2\omega_{\mathbf{q}}}{\hbar}\frac{E_{\mathbf{q},\text{ex-p}}^{\alpha}(\mathbf{r}_{i})\left(E_{\mathbf{q},\text{ex-p}}^{\beta}(\mathbf{r}_{j})\right)^{*}}{\left(\omega\pm i0^{+}\right)^{2}-\omega_{\mathbf{q}}^{2}}.\label{eq:exciton-polariton-GF}
\end{equation}
The above expression assumes that exciton-polaritons have an infinite
lifetime. This is a good approximation at low temperatures, when losses
due to coupling of excitons to phonons are very small \citep{Epstein:2020ab}.
Elastic scattering due to impurities is also small (in Sec. \ref{sec:Conclusions}
we discuss the role of disorder in more detail). In terms of these
Green's functions, the different coefficients in Eq.~\eqref{eq:Lindblad_equation}
are given by
\begin{align}
\Delta_{i} & =\mathcal{P}\int\frac{d\nu}{2\pi}\left[1+2b(\nu)\right]\frac{\mu_{i,\alpha}^{*}A_{\alpha\beta}\left(\nu;\mathbf{r}_{i},\mathbf{r}_{i}\right)\mu_{i,\beta}}{\omega_{0}-\nu},\label{eq:LambShift}\\
g_{ij} & =\mu_{i,\alpha}^{*}D_{\alpha\beta}(\omega_{0};\mathbf{r}_{i},\mathbf{r}_{j})\mu_{j,\beta},\label{eq:coupling}\\
\gamma_{ij} & =\left[1+b(\omega_{0})\right]\mu_{i,\alpha}^{*}A_{\alpha\beta}\left(\omega_{0};\mathbf{r}_{i},\mathbf{r}_{j}\right)\mu_{j,\beta},\label{eq:gamma}\\
\tilde{\gamma}_{ij} & =b(\omega_{0})\mu_{j,\alpha}^{*}A_{\alpha\beta}(\omega_{0};\mathbf{r}_{j},\mathbf{r}_{i})\mu_{i,\beta},\label{eq:dissipation_thermal}
\end{align}
where repeated greek indices are summed over, $\mathcal{P}\int$ denotes
the Cauchy principal value of the integral, and
\begin{equation}
A_{\alpha\beta}\left(\omega;\mathbf{r}_{i},\mathbf{r}_{j}\right)=i\left[D_{\alpha\beta}^{R}(\omega;\mathbf{r}_{i},\mathbf{r}_{j})-D_{\alpha\beta}^{A}(\omega;\mathbf{r}_{i},\mathbf{r}_{j})\right]\label{eq:exciton-polariton_spectral_function}
\end{equation}
is the electric field exciton-polariton spectral function, and
\begin{equation}
D_{\alpha\beta}(\omega;\mathbf{r}_{i},\mathbf{r}_{j})=\frac{1}{2}\left[D_{\alpha\beta}^{R}(\omega;\mathbf{r}_{i},\mathbf{r}_{j})+D_{\alpha\beta}^{A}(\omega;\mathbf{r}_{i},\mathbf{r}_{j})\right]\label{eq:exciton-polartion_Herm_GF}
\end{equation}
is the hermitian part of the Green's function, which can be written
in terms of the spectral function as
\begin{equation}
D_{\alpha\beta}(\omega;\mathbf{r}_{i},\mathbf{r}_{j})=\mathcal{P}\int\frac{d\nu}{2\pi}\frac{A_{\alpha\beta}\left(\nu;\mathbf{r}_{i},\mathbf{r}_{i}\right)}{\omega-\nu}.
\end{equation}
For the rest of this work, we will focus on the zero temperature case.
In that case one has $\lim_{T\rightarrow0}b(\nu)=-\Theta(-\nu)$,
such that $\lim_{T\rightarrow0}\left[1+2b(\nu)\right]=\text{sign}(\nu)$
and $b(\omega_{0})=0$, for $\omega_{0}>0$. Therefore, we conclude
that in the zero temperature limit $\tilde{\gamma}_{ij}=0$.

We notice that the form of Eqs.~\eqref{eq:LambShift}-\eqref{eq:dissipation_thermal}
remains valid if we consider coupling of the NV-centers to the all
the electromagnetic degrees of freedom, instead of only the polariton
mode. In that case, $D_{\alpha\beta}^{R/A}(\omega;\mathbf{r}_{i},\mathbf{r}_{j})$
would be the full electromagnetic Green's function. For a linear medium,
the full electromagnetic Green's function can be obtained from the
classical Maxwell's equations, as the response function to a point
dipole. Replacing the full $D_{\alpha\beta}^{R/A}(\omega;\mathbf{r}_{i},\mathbf{r}_{j})$
by the exciton-polariton contribution amounts to a polariton-pole
approximation to the Green's function, as shown in Appendix~\eqref{appx:polariton-pole}.
This is a good approximation if the exciton-polariton frequency is
close to $\omega_{0}$ and if the NV-centers are in close proximity
to the TMD.

\subsection{Evaluation of the effective couplings and decay rates}

Now that we have obtained both the energy window where polaritons
may be excited and their dispersion relation we can move on to the
explicit calculation of the parameters $\gamma_{ij}$, $\Delta_{i}$
and $g_{ij}$. Starting with $\gamma_{ij}$ and recalling Eqs. (\ref{eq:gamma})
and (\eqref{eq:exciton-polariton-GF}), we obtain

\begin{equation}
\gamma_{ij}=\frac{2\pi}{\hbar}\sum_{\mathbf{q}}\delta\left(\omega_{0}-\omega_{\mathbf{q}}\right)\bm{\mu}_{i}^{*}\cdot\mathbf{E}_{\mathbf{q},\text{ex-p}}(\mathbf{r}_{i})\mathbf{E}_{\mathbf{q},\text{ex-p}}^{*}(\mathbf{r}_{j})\cdot\bm{\mu}_{j},
\end{equation}
Using the previously presented definitions for the mode functions
$\mathbf{E}_{\mathbf{q},\text{ex-p}}(\mathbf{r}_{i})$, converting
the sum into an integral, performing the angular integration, and
finally using the $\delta-$function to compute the remaining integral
we find: 
\begin{multline}
\gamma_{ij}=\frac{2\pi}{\hbar}\frac{q_{0}}{2\pi}\frac{\hbar\omega_{0}}{2\epsilon_{0}N_{\mathbf{q}_{0}}}\left(\frac{\partial\omega_{\mathbf{q}_{0}}}{\partial q}\right)^{-1}\times\\
\times e^{-\kappa_{1,\mathbf{q}}\left(z_{i}+z_{j}\right)}\mathcal{M}_{ij}(\mathbf{q}_{0},\rho),
\end{multline}
where in the reference frame where the dipoles are separated along
the $x$ direction:
\begin{multline}
\mathcal{M}_{ij}(\mathbf{q},\rho)=\mu_{i,x}^{*}\mu_{j,x}\left(J_{0}\left(q\rho\right)-\frac{J_{1}\left(q\rho\right)}{q\rho}\right)\\
+\mu_{i,y}^{*}\mu_{j,y}\frac{J_{1}\left(q\rho\right)}{q\rho}+\mu_{i,z}^{*}\mu_{j,z}\frac{q^{2}}{\kappa_{1,\mathbf{q}}^{2}}J_{0}\left(q\rho\right)\\
+\left(\mu_{i,x}^{*}\mu_{j,z}-\mu_{i,z}^{*}\mu_{j,x}\right)\frac{q}{\kappa_{1,\mathbf{q}}}J_{1}(q\rho).
\end{multline}
For zero separation, one obtains the simpler form
\begin{align}
\mathcal{M}_{ij}(\mathbf{q},0) & =\frac{1}{2}\left(\mu_{i,x}^{*}\mu_{j,x}+\mu_{i,y}^{*}\mu_{j,y}+\frac{2q^{2}}{\kappa_{1,\mathbf{q}}^{2}}\mu_{i,z}^{*}\mu_{j,z}\right).
\end{align}

\begin{figure*}[t]
\centering{}

\includegraphics[width=17.5cm]{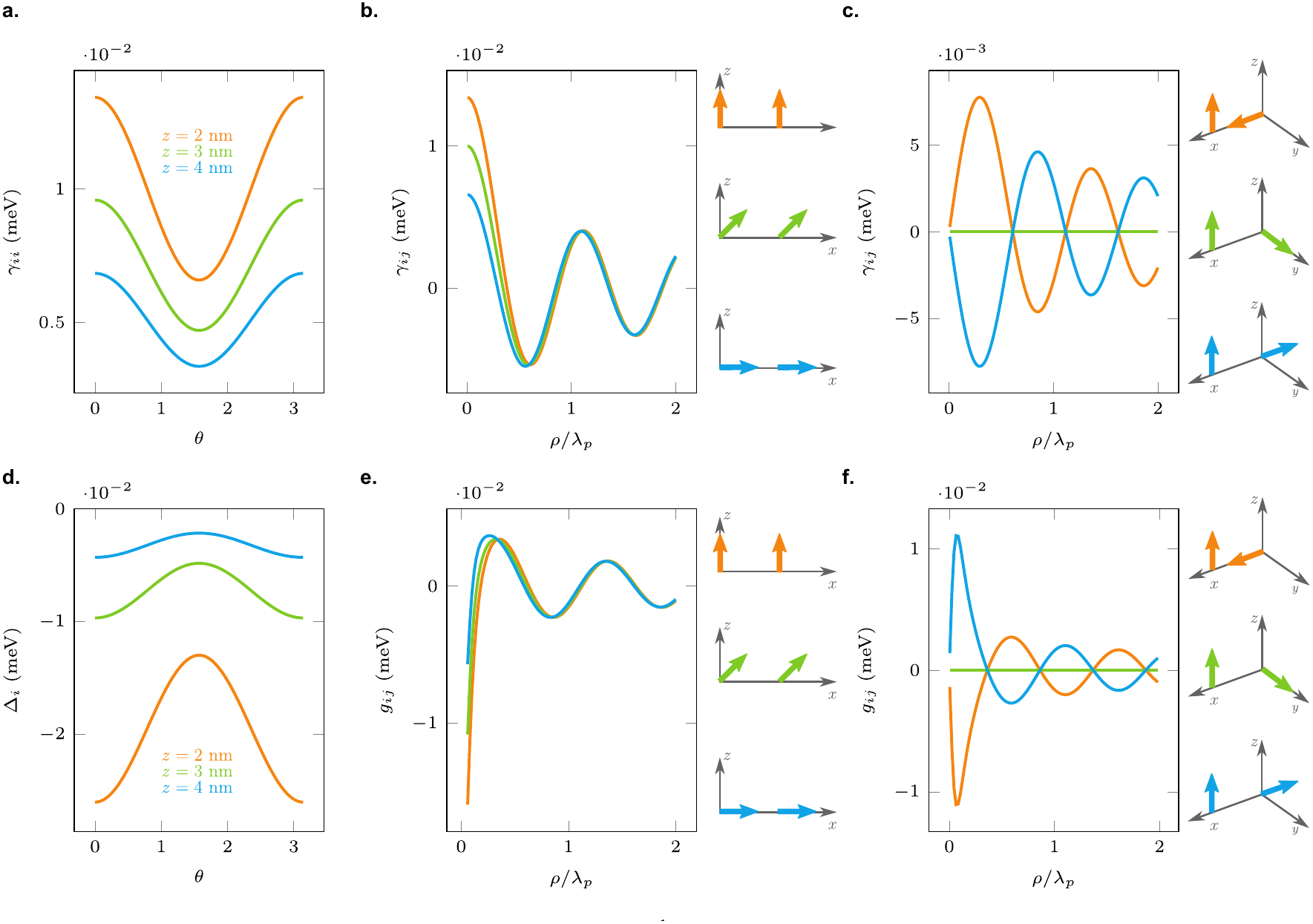}\caption{\label{fig:gamma plots} Interaction parameters, energy shifts, and
damping constants for different configurations of the system. In panel
(a) we depict $\gamma_{11}$ as a function of $\theta$, the angle
the dipole makes with the $z$ axis, for three different NV-center--TMD
separations. In panel (b) we depict $\gamma_{12}$ as a function of
$\rho/\lambda_{p}$, where $\rho$ is the \textcolor{black}{in-plane}
distance between the NV-centers, and $\lambda_{p}$ is the polariton
wavelength for an energy $\hbar\omega_{0}$; in the present case we
have $\lambda_{\text{p}}\approx37$ nm. We consider that both dipoles
lie in the x-z plane, with an angle with the z-axis given by $\theta=0,\,\pi/4$
and $\pi/2$ as depicted in the panel. In panel (c) we show $\gamma_{12}$
as a function of $\rho/\lambda_{p}$, for the configurations shown
in the panel: with one dipole along the $z$ axis while the other
lies in the $x-y$ plane (aligned along $x$, $y$ and $-x$ axes).
In panels (d), (e) and (f) show $\Delta_{i}$ and $g_{12}$ for the
same dipole orientations as in panels (a), (b) and (c). In all plots,
the NV-centers were assumed to be at a distance $z=2$ nm from the
TMD. The parameters of Table \ref{tab:parameters} were used.}
\end{figure*}

To compute the explicit forms of the $\Delta_{i}$ and $g_{ij}$ we
can proceed in a similar way to what we have done with the $\gamma_{ij}$,
the main difference being that we no longer find closed expressions
for these parameters. While for the $\gamma_{ij}$ we had a $\delta-$function
that allowed us to compute the integrals in a completely analytical
way, for the $\Delta_{i}$ and $g_{ij}$ we will have to compute one
of the integrals numerically. Recalling Eqs.~\eqref{eq:coupling}
and \eqref{eq:exciton-polariton-GF}, we obtain: 
\begin{multline}
g_{ij}=\mathcal{P}\int\frac{dqq}{2\pi}e^{-\kappa_{1,\mathbf{q}}\left(z_{i}+z_{j}\right)}\frac{\omega_{\mathbf{q}}}{2\epsilon_{0}N_{\mathbf{q}}}\times\\
\times\left[\frac{1}{\omega_{0}-\omega_{\mathbf{q}}}-\frac{1}{\omega_{0}+\omega_{\mathbf{q}}}\right]\mathcal{M}_{ij}(\mathbf{q},\rho),
\end{multline}
and for the $\Delta_{i}$:
\begin{multline}
\Delta_{i}=\mathcal{P}\int\frac{dqq}{2\pi}e^{-\kappa_{1,\mathbf{q}}\left(z_{i}+z_{j}\right)}\frac{\omega_{\mathbf{q}}}{2\epsilon_{0}N_{\mathbf{q}}}\times\\
\times\left[\frac{1}{\omega_{0}-\omega_{\mathbf{q}}}+\frac{1}{\omega_{0}+\omega_{\mathbf{q}}}\right]\mathcal{M}_{ij}(\mathbf{q},0),
\end{multline}
where $\mathcal{M}_{ij}(\mathbf{q},\rho)$ is defined as before

In Fig. \ref{fig:gamma plots} we present the plot of $\gamma_{ii}$
as a function of $\theta$ and $\gamma_{ij}$ for different dipole
configurations as a function of $\rho$ and $\theta$. In Fig. \ref{fig:gamma plots}
(a) we depict $\gamma_{ii}$ as a function of the angle the dipole
makes with the $z$ axis. We observe that this parameter takes its
minimum value when the dipole is parallel to the TMD plane, and the
maximum value when it is perpendicular to it. Moreover, we also note
the sensitivity of $\gamma_{ii}$ to the dipole--TMD separation.
Small increases on this parameter produce significant changes in the
final result; as the separation to the TMD plane increases the magnitude
of the exciton-polariton induced decay rate diminishes exponentially,
in agreement with the analytical expressions previously found. In
Fig. \ref{fig:gamma plots} (b) we plot $\gamma_{12}$ as a function
of $\rho/\lambda_{\text{p}}$, where $\lambda_{p}=2\pi/q(E_{0})\approx37$
nm, for two parallel dipoles in the $x-z$ plane with different angles
with respect to the $z$-axis. We start by noting that, as expected
from the different elements of $\mathcal{M}_{ij}$, the parameter
$\gamma_{12}$ presents an oscillatory behavior accompanied by a magnitude
decrease as $\rho$ increases. Mathematically, this is a consequence
of the Bessel functions that appear after the angular integration
has been performed leading to a spatial decay proportional to $1/\sqrt{\rho}$
at large distances, where $\rho$ is the distance between NV-centers.
Physically, this is a consequence of the two-dimensional nature of
the polaritons. Furthermore, we observe that $\gamma_{12}$ is bigger
when both dipoles are aligned along the $z$ axis, although the difference
between the different orientations becomes negligible for distances
greater than one polariton wavelength. Finally, in Fig. \ref{fig:gamma plots}
(c) we plot $\gamma_{12}$ as a function of $\rho/\lambda_{\text{p}}$,
only this time we consider one dipole to be aligned along the $z$
axis, while the other is placed on a plane parallel to the $x-y$
plane with different $\phi$ angles. We first note that, just as in
the previous case, $\gamma_{12}$ presents a decaying and oscillatory
behavior as $\rho$ increases. Moreover, we note that $\gamma_{12}$
is symmetric when the dipole is aligned along the positive or negative
$x$ direction, and it vanishes when it is aligned along the $y$
direction, that is, when the dipole is aligned perpendicularly to
the direction connecting the two dipoles.

\textcolor{black}{The plots of $\Delta_{i}$ and $g_{12}$ as a function
of $\theta$, the angle the dipole makes with the $z$ axis, and $\rho$
for different dipole configurations are depicted in Figs. \ref{fig:gamma plots}
(d), (e) and (f). Without surprise, we observe that these quantities
present a similar behavior to $\gamma_{ij}$. There are two aspects
worthy of consideration. (i) Similarly to $\gamma_{12}$ and for not
to small separation, $g_{12}$ decays asymptotically as} as $1/\sqrt{\rho}$,
being roughly in phase opposition to $\gamma_{12}$. (ii) The magnitude
of $\Delta_{i}$ is of the order of 25 $\mu$eV, corresponding to
a small energy renormalization of the NV-centers. Comparing the exciton-polariton
mediated interaction, $g_{12}$ which decays with $\rho^{-1/2}$,
with the electrostatic dipole-dipole interaction, which decays as
$\rho^{-3}$, we conclude that the exciton-polariton interaction dominates
for $\rho\gtrsim\lambda_{\text{p}}/10$.

\section{Exciton mediated superradiance}

\begin{figure}
\includegraphics{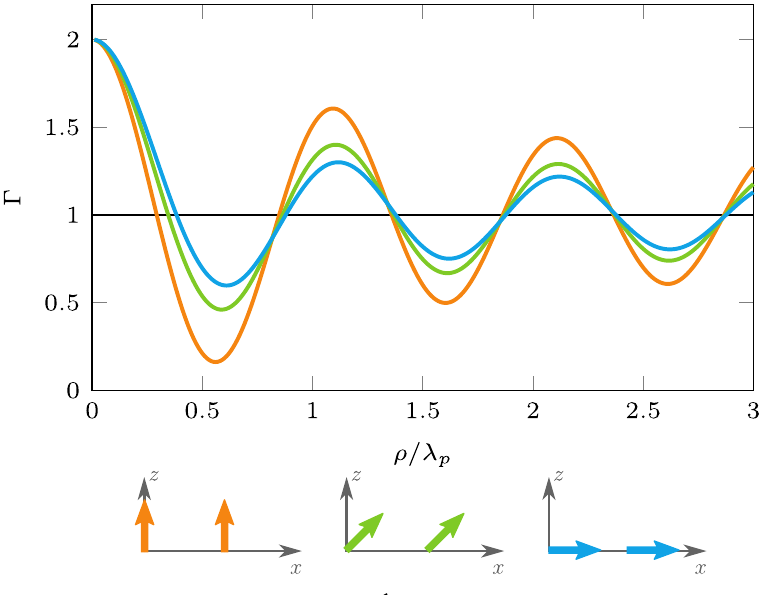}

\caption{\label{fig:Superradiance-in-the}Superradiance and subradiance in
the two-coupled NV-centers system. Three orientations of the two electric
dipole moments relatively $z-$axis are considered. The parameters
are the same as in Fig. \ref{fig:gamma plots}.}
\end{figure}

As noted in Sec. \ref{sec:Hamiltonian}, our model is a modification
of the Dicke Hamiltonian. The modifications are two fold: (i) only
a small number (two) of radiant NV-centers are considered; (ii) there
is more than one bosonic mode, labeled by the in-plane momentum $\mathbf{q}$.
Except for these differences, the model should also present superradiance.
The figure of merit allowing to characterize superradiance is \cite{Huidobro:2012aa}:

\begin{equation}
\Gamma=\frac{\gamma_{11}+\gamma_{12}+\gamma_{21}+\gamma_{22}}{\gamma_{11}+\gamma_{22}}.
\end{equation}
It is known that when $\Gamma>1$ the system shows superradiance (in
the opposite regime, the system is subradiant, that is, the emission
of coherent radiation is suppressed). In Fig. \ref{fig:Superradiance-in-the}
we depict the figure of merit $\Gamma$. It is clear that depending
on the relative position of the two emitters regions exist in space
with $\Gamma>1$. This result lays down the basis for discussing the
$N-$emitters problem.

\section{Conclusions\label{sec:Conclusions}}

In this paper we have described the dynamics of two NV-centers, hosted
by a diamond, coupled to the exciton-polaritons supported by a monolayer
TMD substrate, within the framework of a Lindblad equation. We expressed
the exciton-polarion induce dipole-dipole interaction, energy shifts
and decay rates of the NV-center two-level systems in terms of a Dyadic
Green's function of the exciton-polariton. We have computed this latter
quantity using the modes of the exciton-polaritons alone. We have
found that as a consequence of the two dimensional nature of the exciton-polaritons
the aforementioned parameters decay in space as $\rho^{-1/2}$. This
decay would be stronger with the distance if damping had been considered
thus limiting the range of the interaction between two NV-centers.
Naturally, the intensity of the exponential decay is tied up to the
intensity of the disorder. For small disorder we do not expect an
important effect within distances of the order the exciton-polaritons
wavelength. An estimation of magnitude of disorder on the calculated
parameters can be estimated computing the quantity $f=\exp[-q^{\prime\prime}(\omega)d]$,
where $d$ is the distance between the two NV-centers and $q^{\prime\prime}(\omega)$
is the imaginary part of the exciton-polariton wavenumber. For the
numbers given in Table \ref{tab:parameters} and for $d=\lambda_{\text{p}}/2$
we find $f\simeq0.5$. Therefore, the magnitude of the parameters
represented in the figures would be half of what they are for the
given $d$.

Our results indicate that TMD exciton-polariton mediated super and
subradiance can be observed for NV-centers in diamonds separated by
up to 100 nm. These theoretical predictions can be validated experimentally
by photoluminescent spectroscopy \cite{Scheibner:2007aa}.

Our methods and results are not restricted to NV-centers in diamond,
but can be extended to color centers in other materials, such as quantum
emitters in hexagonal boron nitride (hBN), which have an electric
dipole moment in excess of 2.1 D, that is stronger than for a NV-center
in diamond (smaller values of $\mu\sim0.9-1.1$ D for emitters in
hBN have also been reported \cite{Noh:2018aa,Scavuzzo:2019aa}). Thus,
this work opens the door for the study of quantum optics devices fully
built with van der Waals heterostructures.

\section*{Acknowledgements}

B.A. and N.M.R.P acknowledge support by the Portuguese Foundation
for Science and Technology (FCT) in the framework of the Strategic
Funding UIDB/04650/2020. B.A. further acknowledges support from FCT
through Grant No. CEECIND/02936/2017. N.M.R.P. acknowledges support
from the European Commission through the project ``Graphene-Driven
Revolutions in ICT and Beyond'' (Ref. No. 881603, CORE 3), COMPETE
2020, PORTUGAL 2020, FEDER and the FCT through project POCI-01-0145-FEDER-028114.

\appendix

\section{Details on the derivation of the Lindblad equation\label{appx:Lindblad}}

Here we will provide some details for the derivation of the Linblad
equation. To derive the Lindblad equation that governs the NV-center
degrees of freedom we will treat the TMD exction-polariton field as
a bath which is coupled to the NV-centers via $H_{\text{int}}$ To
study the system's dynamics. The full density matrix of the coupled
NV-centers/polaritons, $\chi(t)$, obeys the following equation in
the interaction picture \citep{Carmichael:2002}: 
\begin{align}
\dot{\chi}_{I}(t) & =-\frac{i}{\hbar}\left[H_{\text{int},I},\chi_{I}(0)\right]\nonumber \\
 & -\frac{1}{\hbar^{2}}\int_{0}^{t}\left[H_{\text{int},I}(t),\left[H_{\text{int},I}(t^{\prime}),\chi_{I}(t^{\prime})\right]\right]dt',\label{eq:d chi_I (t) / dt}
\end{align}
which is easily obtained from the usual equation of motion for density
matrices. The derivation of the Lindblad equation for the density
of matrix for the NV-centers involves a series of hypothesis and approximations.
First, within the Born approximation, it is assumed that at the density
matrix can be written as the product of the density matrix of the
NV-centers, $\rho(t)$, and the time independent density matrix of
the exciton-polariton bath, $R$, i.e. $\chi(t)=\rho(t)\otimes R$.
This hypothesis ignores the initial correlation between the two systems
and assumes that the perturbation to the bath is small. Next, within
the Markov approximation, one replaces $\chi_{I}(t^{\prime})\simeq\rho_{I}(t)\otimes R$
and takes the limit $t\rightarrow\infty$ in the integration in the
second term of Eq.~\eqref{eq:d chi_I (t) / dt}. These simplification
is justified if the characteristic time of the bath is much shorter
than the characteristic time of the system, which leads to a loss
of memory and Markovian behavior. Finally, within the post-trace rotating
wave approximation, only energy conserving terms are kept. Applying
these approximations to Eq. (\ref{eq:d chi_I (t) / dt}) and tracing
out the degrees of freedom of the exciton-polariton bath, and returning
to the Schrödinger picture, the equation of motion for the NV-center
density matrix is given by \cite{BreuerPetruccione}

\begin{widetext}

\begin{align}
\frac{\partial\rho(t)}{\partial t}= & -\frac{i}{\hbar}\left[\sum_{i}\frac{1}{2}\hbar\omega_{0}\sigma_{i}^{z}+\sum_{i,j}\sum_{s=\pm}\mu_{i,\alpha}^{s}S_{\alpha\beta}(s\omega_{0};\mathbf{r}_{i},\mathbf{r}_{j})\mu_{j,\beta}^{\bar{s}}\sigma_{i}^{s}\sigma_{j}^{\bar{s}},\rho_{I}(t)\right]\nonumber \\
 & +\frac{1}{\hbar}\sum_{i,j}\sum_{s=\pm}\mu_{i,\alpha}^{s}\gamma_{\alpha\beta}(s\omega_{0};\mathbf{r}_{i},\mathbf{r}_{j})\mu_{j,\beta}^{\bar{s}}\left(\sigma_{j}^{\bar{s}}\rho_{I}^{S}(t)\sigma_{i}^{s}-\frac{1}{2}\left\{ \sigma_{i}^{s}\sigma_{j}^{\bar{s}},\rho_{I}^{S}(t)\right\} \right),
\end{align}
where $\mu_{i,\alpha}^{+}=\mu_{i,\alpha}^{*}$, $\mu_{i,\alpha}^{-}=\mu_{i,\alpha}$,
$\bar{s}=-s$, and
\begin{align}
S_{\alpha\beta}(\omega;\mathbf{r}_{i},\mathbf{r}_{j}) & =P\int\frac{d\nu}{2\pi}\left[1+b(\nu)\right]\frac{A_{\alpha\beta}\left(\nu;\mathbf{r}_{i},\mathbf{r}_{j}\right)}{\omega-\nu},\\
\gamma_{\alpha\beta}(\omega;\mathbf{r}_{i},\mathbf{r}_{j}) & =\left[1+b(\omega)\right]A_{\alpha\beta}\left(\omega;\mathbf{r}_{i},\mathbf{r}_{j}\right),
\end{align}
with $A_{\alpha\beta}\left(\nu;\mathbf{r}_{i},\mathbf{r}_{j}\right)$
the electric field spectral function. Using the fact that $A_{\alpha\beta}\left(\omega;\mathbf{r}_{i},\mathbf{r}_{j}\right)=-A_{\beta\alpha}\left(-\omega;\mathbf{r}_{j},\mathbf{r}_{i}\right)$,
it is easy to see that the dissipators in the main text are given
by
\begin{align}
\gamma_{ij} & =\mu_{i,\alpha}^{+}\gamma_{\alpha\beta}(\omega_{0};\mathbf{r}_{i},\mathbf{r}_{j})\mu_{j,\beta}^{-},\\
\tilde{\gamma}_{ij} & =\mu_{i,\alpha}^{-}\gamma_{\alpha\beta}(-\omega_{0};\mathbf{r}_{i},\mathbf{r}_{j})\mu_{j,\beta}^{+}.
\end{align}

We will rewrite the terms involving $S_{\alpha\beta}(s\omega_{0};\mathbf{r}_{i},\mathbf{r}_{j})$
by separating the terms with $i\neq j$ and $i=j$. For the term with
$i\neq j$, we write
\begin{multline}
\sum_{i\neq j}\sum_{s=\pm}\mu_{i,\alpha}^{s}S_{\alpha\beta}(s\omega_{0};\mathbf{r}_{i},\mathbf{r}_{j})\mu_{j,\beta}^{\bar{s}}\sigma_{i}^{s}\sigma_{j}^{\bar{s}}=\\
=\sum_{i\neq j}\mu_{i,\alpha}^{+}S_{\alpha\beta}(\omega_{0};\mathbf{r}_{i},\mathbf{r}_{j})\mu_{j,\beta}^{-}\sigma_{i}^{+}\sigma_{j}^{-}+\mu_{i,\alpha}^{-}S_{\alpha\beta}(-\omega_{0};\mathbf{r}_{i},\mathbf{r}_{j})\mu_{j,\beta}^{+}\sigma_{i}^{-}\sigma_{j}^{+}\\
=\sum_{i\neq j}\mu_{i,\alpha}^{+}S_{\alpha\beta}(\omega_{0};\mathbf{r}_{i},\mathbf{r}_{j})\mu_{j,\beta}^{-}\sigma_{i}^{+}\sigma_{j}^{-}+\mu_{j,\beta}^{-}S_{\beta\alpha}(-\omega_{0};\mathbf{r}_{j},\mathbf{r}_{i})\mu_{i,\alpha}^{+}\sigma_{j}^{-}\sigma_{i}^{+}.
\end{multline}
Using the fact that $\left[\sigma_{i}^{+},\sigma_{j}^{-}\right]=0$
for $i\neq j$, we can write
\begin{equation}
\sum_{i\neq j}\sum_{s=\pm}\mu_{i,\alpha}^{s}S_{\alpha\beta}(s\omega_{0};\mathbf{r}_{i},\mathbf{r}_{j})\mu_{j,\beta}^{\bar{s}}\sigma_{i}^{s}\sigma_{j}^{\bar{s}}=\sum_{i\neq j}g_{ij}\sigma_{i}^{+}\sigma_{j}^{-}.
\end{equation}
where 
\begin{align}
g_{ij} & =\mu_{i,\alpha}^{+}\left[S_{\alpha\beta}(\omega_{0};\mathbf{r}_{i},\mathbf{r}_{j})+S_{\beta\alpha}(-\omega_{0};\mathbf{r}_{j},\mathbf{r}_{i})\right]\mu_{j,\beta}^{-}\nonumber \\
 & =P\int\frac{d\nu}{2\pi}\frac{\mu_{i,\alpha}^{+}A_{\alpha\beta}\left(\nu;\mathbf{r}_{i},\mathbf{r}_{j}\right)\mu_{j,\beta}^{-}}{\omega_{0}-\nu}\nonumber \\
 & =\mu_{i,\alpha}^{+}D_{\alpha\beta}\left(\omega_{0};\mathbf{r}_{i},\mathbf{r}_{j}\right)\mu_{j,\beta}^{-}.
\end{align}
For the term with $i=j$ we write
\begin{equation}
\sum_{i}\sum_{s=\pm}\mu_{i,\alpha}^{s}S_{\alpha\beta}(s\omega_{0};\mathbf{r}_{i},\mathbf{r}_{i})\mu_{j,\beta}^{\bar{s}}\sigma_{i}^{s}\sigma_{i}^{\bar{s}}=\sum_{i}\mu_{i,\alpha}^{+}S_{\alpha\beta}(\omega_{0};\mathbf{r}_{i},\mathbf{r}_{i})\mu_{i,\beta}^{-}\sigma_{i}^{+}\sigma_{i}^{-}+\mu_{i,\alpha}^{-}S_{\alpha\beta}(-\omega_{0};\mathbf{r}_{i},\mathbf{r}_{i})\mu_{i,\beta}^{+}\sigma_{i}^{-}\sigma_{i}^{+}.
\end{equation}
Noticing that
\begin{align}
\sigma_{i}^{+}\sigma_{i}^{-} & =\left|e,i\right\rangle \left\langle e,i\right|\\
\sigma_{i}^{-}\sigma_{i}^{+} & =\left|g,i\right\rangle \left\langle g,i\right|
\end{align}
We can then write
\begin{equation}
\sum_{i}\sum_{s=\pm}\mu_{i,\alpha}^{s}S_{\alpha\beta}(s\omega_{0};\mathbf{r}_{i},\mathbf{r}_{i})\mu_{j,\beta}^{\bar{s}}\sigma_{i}^{s}\sigma_{i}^{\bar{s}}=\frac{1}{2}\tilde{\Delta}_{i}\sigma_{i}^{0}+\frac{1}{2}\Delta_{i}\sigma_{i}^{z},
\end{equation}
where $\sigma_{i}^{0}=\left|e,i\right\rangle \left\langle e,i\right|+\left|g,i\right\rangle \left\langle g,i\right|$,
\begin{align}
\Delta_{i} & =\mu_{i,\alpha}^{+}\left[S_{\alpha\beta}(\omega_{0};\mathbf{r}_{i},\mathbf{r}_{i})-S_{\beta\alpha}(-\omega_{0};\mathbf{r}_{i},\mathbf{r}_{i})\right]\mu_{i,\beta}^{-}\nonumber \\
 & =P\int\frac{d\nu}{2\pi}\left[1+2b(\nu)\right]\frac{\mu_{i,\alpha}^{+}A_{\alpha\beta}\left(\nu;\mathbf{r}_{i},\mathbf{r}_{i}\right)\mu_{i,\beta}^{-}}{\omega_{0}-\nu},
\end{align}
and
\begin{align}
\tilde{\Delta}_{i} & =\mu_{i,\alpha}^{+}\left[S_{\alpha\beta}(\omega_{0};\mathbf{r}_{i},\mathbf{r}_{i})+S_{\beta\alpha}(-\omega_{0};\mathbf{r}_{i},\mathbf{r}_{i})\right]\mu_{i,\beta}^{-}\nonumber \\
 & =P\int\frac{d\nu}{2\pi}\frac{\mu_{i,\alpha}^{+}A_{\alpha\beta}\left(\nu;\mathbf{r}_{i},\mathbf{r}_{i}\right)\mu_{i,\beta}^{-}}{\omega_{0}-\nu}\nonumber \\
 & =\mu_{i,\alpha}^{+}D_{\alpha\beta}\left(\omega_{0};\mathbf{r}_{i},\mathbf{r}_{i}\right)\mu_{i,\beta}^{-}.
\end{align}
If the NV-centers are equal, we have $\tilde{\Delta}_{1}=\tilde{\Delta}_{2}$,
and therefore, this term only leads to a global shift in energy, which
can therefore be ignored.

\end{widetext}

\section{Mode length and Green's function\label{appx:Mode-length-and}}

In this appendix we give further insight on the derivation of the
mode length and Green function presented in the main text.

Following Ref. \cite{Ferreira:2020aa}, the mode length is defined
as:
\begin{equation}
N(\mathbf{q})=\int\mathbf{A}_{\mathbf{q}}^{*}(z)\cdot\left(\bar{\epsilon}_{r}(z)+\frac{\omega}{2}\frac{\partial}{\partial\omega}\bar{\epsilon}_{r}(z)\right)\cdot\mathbf{A}_{\mathbf{q}}(z)dz,
\end{equation}
where:
\begin{equation}
\epsilon_{r}(z)=\bar{\epsilon}_{d}(z)+\frac{i\bar{\sigma}(\omega)}{\epsilon_{0}\omega}\delta(z),
\end{equation}
with\textcolor{red}{{} }$\bar{\epsilon}_{d}(z)$ the dielectric function
tensor of the different layers composing the system, $\bar{\sigma}(\omega)$
the conductivity tensor of the TMD and:
\[
\mathbf{A}_{\mathbf{q}}(z)=\begin{cases}
\mathbf{u}_{1,\mathbf{q}}^{+}e^{-\kappa_{1,\mathbf{q}}z} & z>0\\
\mathbf{u}_{2,\mathbf{q}}^{-}e^{\kappa_{2,\mathbf{q}}z} & z<0
\end{cases}
\]
where:
\begin{equation}
\mathbf{u}_{n,\mathbf{q}}^{\pm}=i\frac{\mathbf{q}}{q}\mp\frac{q}{\kappa_{n,\mathbf{q}}}\hat{\mathbf{z}}.
\end{equation}
Noting that the conductivity tensor only couples to the in-plane degrees
of freedom, and observing that the exciton conductivity can be written
as:
\begin{equation}
\frac{\sigma(\omega)}{\epsilon_{0}\omega}=-id\chi_{bg}+i\frac{df_{\text{ex}}}{2}\frac{\omega_{\text{ex}}^{2}}{\omega^{2}-\omega_{ex}^{2}},
\end{equation}
where the imaginary part of the susceptibility $\chi(\omega)$ was
discarded, we find:
\begin{multline}
N(\mathbf{q})=\frac{\epsilon_{1}}{2}\frac{\kappa_{1,\mathbf{q}}^{2}+q^{2}}{\kappa_{1,\mathbf{q}}^{3}}+\frac{\epsilon_{2}}{2}\frac{\kappa_{2,\mathbf{q}}^{2}+q^{2}}{\kappa_{2,\mathbf{q}}^{3}}+\frac{i\sigma(\omega)}{2\epsilon_{0}\omega}+\frac{i}{2\epsilon_{0}}\frac{\partial\sigma(\omega)}{\partial\omega}\\
=\frac{\epsilon_{1}}{2}\frac{\kappa_{1,\mathbf{q}}^{2}+q^{2}}{\kappa_{1,\mathbf{q}}^{3}}+\frac{\epsilon_{2}}{2}\frac{\kappa_{2,\mathbf{q}}^{2}+q^{2}}{\kappa_{2,\mathbf{q}}^{3}}+\\
+\frac{1}{2}df_{\text{ex}}\frac{\omega_{\text{ex}}^{4}}{\left(\omega_{\mathbf{q}}^{2}-\omega_{\text{ex}}^{2}\right)^{2}}+d\chi_{bg}.
\end{multline}
The vector potential operator for the exciton-polariton is therefore,
written as
\begin{equation}
\mathbf{A}_{\text{ex-p}}(\mathbf{r})=\sum_{\mathbf{q}}\left(\mathbf{A}_{\mathbf{q},\text{ex-p}}(\mathbf{r})a_{\mathbf{q}}+\mathbf{A}_{\mathbf{q},\text{ex-p}}^{*}(\mathbf{r})a_{\mathbf{q}}^{\dagger}\right)
\end{equation}
where
\begin{multline}
\mathbf{A}_{\mathbf{q},\text{ex-p}}(\mathbf{r})=\sqrt{\frac{\hbar}{2A\epsilon_{0}\omega_{\mathbf{q}}N(\mathbf{q})}}\times\\
\times e^{-\kappa_{1,\mathbf{q}}z}e^{i\mathbf{q}\cdot\mathbf{x}}\begin{cases}
\left(i\frac{\mathbf{q}}{q}-\frac{q}{\kappa_{1,\mathbf{q}}}\hat{\mathbf{z}}\right)e^{-\kappa_{1,\mathbf{q}}z}, & z>0\\
\left(i\frac{\mathbf{q}}{q}+\frac{q}{\kappa_{2,\mathbf{q}}}\hat{\mathbf{z}}\right)e^{\kappa_{2,\mathbf{q}}z}, & z<0
\end{cases}
\end{multline}
The electric field operator for the exciton-polariton is then given
by Eq.~\eqref{eq:E_expansion}, with $\mathbf{E}_{\mathbf{q},\text{ex-p}}(\mathbf{r})=i\omega_{\mathbf{q}}\mathbf{A}_{\mathbf{q},\text{ex-p}}(\mathbf{r})$.

\section{Exciton-polariton Green's function as a polariton-pole approximation\label{appx:polariton-pole}}

In this appendix we will show how the exciton-polariton Green's function
emerges as a polariton-pole approximation to the full electric field
Green's function. For the structure considered in this work, the Green's
function can be written as (for $z_{i},z_{j}>0$):
\begin{multline}
\bm{D}^{R}(\omega;\mathbf{r}_{i},\mathbf{r}_{j})=\bm{D}^{R,0}(\omega;\mathbf{r}_{i},\mathbf{r}_{j})\\
+\frac{1}{A}\sum_{\mathbf{q},\lambda}e^{i\mathbf{q}\cdot\left(\mathbf{x}_{i}-\mathbf{x}_{j}\right)}\bm{D}_{\lambda}^{R}(\omega;\mathbf{q},z_{i},z_{j}),
\end{multline}
where $\bm{D}^{R,0}(\omega;\mathbf{r}_{i},\mathbf{r}_{j})$ is the
Green's function for the field in vacuum and $\bm{D}_{\lambda}^{R}(\omega;\mathbf{q},z_{i},z_{j})$
is the reflected Green's function for the $\lambda=s,p$ polarization.
These are given by
\begin{equation}
\bm{D}_{\lambda}^{R}(\omega;\mathbf{q}_{i},z_{i},z_{j})=-\mu_{0}\omega^{2}\frac{i}{2k_{z,1}}r_{\lambda}e^{ik_{z,1}\left(z_{i}+z_{j}\right)}\mathbf{e}_{\lambda,1}^{+}\otimes\mathbf{e}_{\lambda,1}^{-},
\end{equation}
where $k_{z,n}=i\kappa_{n}$, $\mathbf{e}_{\lambda,n}^{\pm}$ are
polarization vectors and $r_{\lambda}$ reflection coefficients for
the $\lambda=s,p$ polarizations. Focusing on the $p$-polarization,
we have
\begin{align}
\mathbf{e}_{p,n}^{\pm} & =\frac{q}{k_{n}}\mathbf{e}_{z}\mp\frac{k_{z,n}}{k_{n}}\frac{\mathbf{q}}{q},\\
r_{p} & =\frac{\frac{\omega^{2}\epsilon_{2}}{\kappa_{2}}-\frac{\omega^{2}\epsilon_{1}}{\kappa_{1}}+\frac{i}{\epsilon_{0}}\omega\sigma(\omega)}{\frac{\omega^{2}\epsilon_{2}}{\kappa_{2}}+\frac{\omega^{2}\epsilon_{1}}{\kappa_{1}}+\frac{i}{\epsilon_{0}}\omega\sigma(\omega)}
\end{align}
with $k_{n}^{2}=\epsilon_{1}\omega^{2}/c^{2}$. At the exciton-polariton
dispersion relation, $r_{p}$ has a pole. Let us expand the Green's
function around this pole. Defining
\begin{equation}
d_{\mathbf{q}}(\omega)=\frac{\omega^{2}\epsilon_{2}}{\kappa_{z,2}}+\frac{\omega^{2}\epsilon_{1}}{\kappa_{z,1}}+\frac{i}{\epsilon_{0}}\omega\sigma(\omega),
\end{equation}
the exciton-polariton dispersion relation is defined by $d_{\mathbf{q}}(\omega_{\mathbf{q}})=0$.
Expanding $d_{\mathbf{q}}(\omega)$ around $\omega_{\mathbf{q}}$
and keeping only the imaginary part of $\sigma(\omega)$, we obtain
\begin{equation}
d_{\mathbf{q}}(\omega)\simeq\left.\frac{\partial d_{\mathbf{q}}(\omega)}{\partial\left(\omega^{2}\right)}\right|_{\omega=\omega_{\mathbf{q}}}\left(\omega^{2}-\omega_{\mathbf{q}}^{2}\right),
\end{equation}
where
\begin{multline}
\left.\frac{\partial d_{\mathbf{q}}(\omega)}{\partial\left(\omega^{2}\right)}\right|_{\omega=\omega_{\mathbf{q}}}=\left.\frac{\epsilon_{1}\left(q^{2}+\kappa_{z,1}^{2}\right)}{2\kappa_{z,1}^{3}}\right|_{\omega=\omega_{\mathbf{q}}}+\\
\left.\frac{\epsilon_{2}\left(q^{2}+\kappa_{z,2}^{2}\right)}{2\kappa_{z,2}^{3}}\right|_{\omega=\omega_{\mathbf{q}}}+\frac{i}{\epsilon_{0}}\left.\frac{\partial\left(\omega\sigma(\omega)\right)}{\partial\left(\omega^{2}\right)}\right|_{\omega=\omega_{\mathbf{q}}},
\end{multline}
which we recognize to be nothing, but the mode length, $N_{\mathbf{q}}$.
Therefore, we obtain the polariton-pole contribution to reflection
coefficient
\begin{equation}
r_{p}\simeq\left[-\frac{2\omega_{\mathbf{q}}^{2}\epsilon_{1}}{\kappa_{z,1}}\right]_{\omega=\omega_{p}}\frac{1}{N_{\mathbf{q}}}\frac{1}{\omega^{2}-\omega_{\mathbf{q}}^{2}}
\end{equation}
approximating the full Green's function by its polariton-pole contribution,
we obtain
\begin{multline}
\bm{D}^{R}(\omega;\mathbf{r}_{i},\mathbf{r}_{j})\simeq\frac{1}{A}\sum_{\mathbf{q},\lambda}e^{i\mathbf{q}\cdot\left(\mathbf{x}_{i}-\mathbf{x}_{j}\right)}e^{-\kappa_{1}\left(z_{i}+z_{j}\right)}\frac{2\omega_{\mathbf{q}}}{\omega^{2}-\omega_{\mathbf{q}}^{2}}\\
\frac{\omega_{\mathbf{q}}}{2\epsilon_{0}N_{\mathbf{q}}}\left[i\frac{\mathbf{q}}{q}-\frac{q}{\kappa_{1,\mathbf{q}}}\mathbf{e}_{z}\right]\otimes\left[-i\frac{\mathbf{q}}{q}-\frac{q}{\kappa_{1,\mathbf{q}}}\mathbf{e}_{z}\right],
\end{multline}
which coincides with Eq.~\ref{eq:exciton-polariton-GF} of the main
text.


%

\end{document}